\documentclass[10pt]{article}
    \usepackage{amsmath,amssymb,bm,epsf,epsfig,graphicx}
    \input epsf.sty
    \usepackage[dvipsnames]{xcolor}
    \usepackage[utf8]{inputenc}
    \usepackage[english]{babel}
    \usepackage{amsmath}
    \usepackage{latexsym}
    \usepackage{amsfonts}
    \usepackage{mathtools}
    \usepackage{amssymb}
    \usepackage{scalerel}
    \usepackage{extpfeil}
    \usepackage[left=3cm,right=3cm,top=3.1cm,bottom=3.1cm]{geometry}
    \usepackage{cite}
    \usepackage{tensor}
    \usepackage{tikz}
    \usepackage{tikz-cd}
    \usepackage[mathscr]{euscript}
    \usetikzlibrary{arrows.meta, bending}
    \tikzset{C/.style={circle, minimum size=8mm,
    		node contents={},
    		append after command={\pgfextra{%
    				\draw[-{Straight Barb[flex']}](\tikzlastnode.150) arc (450:110:2.8mm);}
    	}}
    }
    
    \newcommand{\bs}[1]{\boldsymbol{#1}}
    
    \usepackage{hyperref}
    \numberwithin{equation}{section}

    \def\p{\partial}

    \def \be {\begin{eqnarray}}
    \def \ee {\end{eqnarray}}
    \def \bal {\begin{align}}
    \def \eal {\end{align}}
    \def \bdm {\begin{displaymath}}
    \def \edm {\end{displaymath}}

    \def\0{\nonumber}

    \def\wc{\omega_\text{c}}
    \def\wo{\omega_\text{o}}
    \def\hw{\hat\omega}

    \def\bfm{\boldsymbol{m}}
    \def\bfl{\boldsymbol{l}}
    \def\bfn{\boldsymbol{n}}
    \def\bdel{\boldsymbol{\partial}}
    \def \bfU{\boldsymbol{U}}

    \newcommand{\ch}[1]{{\color{black}{#1}}}
    
\begin{document}
    
    	\begingroup\allowdisplaybreaks
    
    \vspace*{1.1cm}
    
    \centerline{\Large \bf  The Nilpotent Structure}
    \vspace {.3cm}
    
    \centerline{\Large \bf of Open-Closed String Field Theory  } 
    
    \vspace{.3cm}

    \begin{center}

    {\large Carlo Maccaferri$^{(a)}$\footnote{Email: maccafer at gmail.com}, Alberto Ruffino$^{(a)}$\footnote{Email: ruffinoalb at gmail.com}  and  Jakub Vo\v{s}mera$^{(b)}$\footnote{Email: jvosmera at phys.ethz.ch} }
    \vskip 1 cm
    $^{(a)}${\it Dipartimento di Fisica, Universit\`a di Torino, \\INFN  Sezione di Torino \\
    Via Pietro Giuria 1, I-10125 Torino, Italy}
    \vskip .5 cm
    \vskip .5 cm
    $^{(b)}${\it Institut f\"{u}r Theoretische Physik, ETH Z\"{u}rich\\
    	Wolfgang-Pauli-Straße 27, 8093 Z\"{u}rich, Switzerland}
    
    \end{center}
    
    \vspace*{6.0ex}
    
    \centerline{\bf Abstract}
    \bigskip
    In this note we revisit the homotopy-algebraic structure of oriented bosonic open-closed string field theory and we give a new compact formulation in terms of a single cyclic open-closed co-derivation which defines a single nilpotent structure describing the consistency of generic open-closed off-shell amplitudes with arbitrary number of boundaries and at arbitrary genus. 
        \baselineskip=16pt
    \newpage
    \tableofcontents

    \section{Introduction and summary}\label{sec:1}
    
   Open-closed string field theory is a complete QFT framework to address physical questions in string theory when effects involving D-branes are important. The structure of the theory is deeply linked to the structure of the possible sewings of Riemann surfaces with boundaries and its quantum consistency has been elucidated by Zwiebach \cite{Zwiebach:1990qj, Zwiebach:1997fe} in the framework of Batalin-Vilkovisky (BV) quantization. Truncating the theory to only punctured spheres and disks, an interesting mathematical structure called Open-Closed Homotopy Algebra (OCHA) was extracted by Kajiura and Stasheff in \cite{Kajiura:2004xu, Kajiura:2005sn}\footnote{Kajiura-Stasheff homotopy structure has been recently generalized to the RNS superstring in \cite{Kunitomo:2022qqp}.} and recently  extended by two of us \cite{cosmo} to a full Sphere-Disk Homotopy Algebra (SDHA) by also including couplings with only closed strings on the disk.  This mathematical structure turned out to be useful to describe classical open string field theory in a deformed background given by a classical closed string field theory solution, with frozen closed string degrees of freedom. In particular the SDHA extension was shown to be crucial to correctly capture gauge-invariant observables.
   
Because of the familiar open-closed channel duality relating an open string quantum loop to a classical closed string exchange, there is not a classical open-closed SFT involving propagating open and closed strings:  the theory is intrinsically quantum mechanical. The homotopy-algebraic structure at the quantum level turned out to be rather involved (certainly more involved  than quantum closed string field theory \cite{Zwiebach:1992ie}, see for example \cite{Erler:2019loq}) and has been addressed in \cite{Munster:2011ij} in terms of open and closed $IBL_\infty$ algebras and morphisms between the two, in a structure called Quantum Open-Closed Homotopy Algebra. In this approach, purely open and purely closed  couplings are described by (generalized) co-derivations and the other open-closed couplings are associated to morphisms intertwining between purely closed and purely open interactions. Beside the mathematical interest, this structure treats open-closed interactions on a different level than purely open or purely closed ones and therefore it does not seem very convenient to work with in concrete settings when one would like to treat all path integral variables, and their interactions, on the same footing. Along this direction, a more conservative and conceptually economic construction has been provided  recently in \cite{Moosavian:2019ydz}, where all open-closed couplings have been associated to a common set of open-closed multistring products, thus providing a picture which is closer to  what happens  in a standard QFT when different fields are present. However open strings there have been treated fully symmetrically and the notion of colour ordering (together with  other details) has been lost in the construction. 
   
   The aim of this note is to provide a new formulation where, as the physical intuition suggests, closed strings are taken fully symmetrically and open strings are taken to be cyclic on every boundary, as in the setting of \cite{Zwiebach:1997fe, Munster:2011ij} but, at the same time, all open-closed interactions are unified inside a single object, in a way that is more similar in spirit to \cite{Moosavian:2019ydz}.  As this is evidently going to be rather technical,  let us summarize the main outcomes of the paper. 
    
    The full (bosonic) open-closed quantum BV master action can be written in WZW form as
    \begin{align}
    S_{\rm oc}[\Psi,\Phi]=\int_0^1dt\, \,\hat\omega\left(\dot\chi(t)\,,\,\pi_1\,{\bfn}\,{\cal G}(t)\right),\label{compaction}
    \end{align}
    where $\Psi$ is the open string field, $\Phi$ is the closed and $\chi(t)=\Phi(t)+\Psi(t)$ is the (interpolated) open-closed string field.
    The various objects entering the above expression are defined in the body of the paper but anticipating without too many details , $\hat \omega$ is a symplectic form in the space ${\cal H}_{\rm closed}\oplus {\cal H}_{\rm open}$ where the dynamical closed-open BV string field  $\chi$ lives. Specifically
    \begin{align}
    \hw(\chi_1\,,\,\chi_2)\coloneqq\frac\wc{\kappa^2}(\Phi_1,\Phi_2)+\frac\wo\kappa(\Psi_1,\Psi_2),
    \end{align} 
    where $\kappa$ is the string coupling constant.
    ${\cal G}(t)$ is the (interpolated) group-like element of the open-closed tensor algebra ${\cal SH}_{\rm c}\otimes' {\cal SCH}_{\rm o}$ where closed strings are fully symmetrized and open strings are cyclic on every boundary and every boundary is fully symmetrized with the others. The open-closed string field $\chi$ can be extracted from ${\cal G}$ by taking the projection onto a single (open or closed) string 
    as $\chi=\pi_1 \,{\cal G}$. Finally ${\bfn}$ is an {\color{Black} odd} open-closed co-derivation which is cyclic with respect to the open-closed symplectic form $\hat\omega$ (a notion which will be discussed in the body of the paper). The co-derivation $\bfn$ can be decomposed according to the closed or open string output 
   \be
   \bfn=\bfl+\bfm,
   \ee
   where both $\bfl$ and $\bfm$ are  sums of co-derivations associated to genus $g$ Riemann surfaces with $b$ boundaries and arbitrary number of open and closed insertions
\begin{align}
{\bfl}&=\sum_{g,b}\kappa^{2g+b}{\bfl}^{(g,b)},\\
{\bfm}&=\sum_{g,b}\kappa^{2g+b-1}{\bfm}^{(g,b)}.
\end{align}
\textcolor{Black}{The difference in the two objects is that $\boldsymbol{l}$ has a closed string output while $\boldsymbol{m}$ has an open string output. Calling $\pi_{10}$ and $\pi_{01}$ the projections on respectively one single closed or open string, this means
\begin{align}
\pi_{01}\boldsymbol{l}=0,\\
\pi_{10}\boldsymbol{m}=0.
\end{align}
}\\
   {\color{Black}The compact form of the action \eqref{compaction} can be unpackaged using the definitions in the paper to any desired order. For example we can make explicit the kinetic terms and the first few sphere and disk couplings as   
 \begin{equation}
  \begin{split}
       S_{\rm oc}[\Psi,\Phi]&=\frac{1}{2\kappa^2}\wc(\Phi,Q_{\rm c}\Phi)+\frac{1}{3!\kappa^2}\wc\left(\Phi,l_{2,0}^{(0,0)}(\Phi,\Phi)\right)+\cdots\\
       &+\frac{1}{2\kappa}\wo(\Psi,Q_{\rm o}\Psi)+\frac{1}{3\kappa}\wo\left(\Psi,m_{0,2}^{(0,1)}(\Psi,\Psi)\right)+\cdots\\
       &+\frac{1}{\kappa}\wc\left(\Phi,l_{0,0}^{(0,1)}\right)+\frac{1}{2!\kappa}\wc\left(\Phi,l_{1,0}^{(0,1)}(\Phi)\right)+\cdots\\
       & +\frac{1}{\kappa}\wo\left(\Psi,m_{1,0}^{(0,1)}(\Phi)\right)+\cdots,
       \end{split}
  \end{equation}
  where the upper indices $\cdot^{(g,b)}$ denote genus and boundaries, while the lower indices $\cdot_{n_{\rm c},n_{\rm o}}$denote the number of closed and open-string inputs (in case of a single boundary).
The first two lines are the purely closed and open parts of the action respectively on the sphere and on the disk, the third line are closed string couplings on the disk without explicit open string insertions (notice in particular the boundary state $l_{0,0}^{(0,1)}$ which acts as a zero-product / tadpole for the closed strings). Finally the last term corresponds to the coupling between an open and a closed string on the disk. Notice that this coupling can be written in two different ways using the  relation \eqref{dual}
\begin{equation}
    \wo\left(\Psi,m_{1,0}^{(0,1)}(\Phi)\right)=(-)^{d(\Phi)+d(\Psi)+d(\Phi)d(\Psi)}\wc\left(\Phi,l_{0,1}^{(0,1)}(\Psi)\right).
\end{equation}
This possibility of  defining the same coupling using a product with a closed or an open string output is present every time there is at least one open and one closed string in interaction. Therefore  $\bfl$ and $\bfm$ are not independent. Indeed, at the open-closed  co-algebra level, they are related  by the open-closed duality relations (\ref{opclcycl1}, \ref{opclcycl2}), making explicit the very fact that any open-closed correlator can be computed by factorization on either closed or open string channels (on every possible boundary with at least an open string puncture). \\

  }
  With these premises, the consistency of the action is provided by the BV quantum master equation which can be explicitly evaluated to
    \begin{align}
    \frac12\left(S_{\rm oc},S_{\rm oc}\right)+\Delta S_{\rm oc}=\int_0^1dt\,\, \hat\omega\left(\dot\chi\,,\,\pi_1\left({\bfn}+\bfU\right)^2\,{\cal G}(t)\right),
    \end{align}
    where $\bfU$ is the higher order co-derivation associated to the Poisson bivector $ U$ defining the inverse of $\hat\omega$ in the standard sense that $(\hat \omega\otimes \boldsymbol{1})(1\otimes U)=\boldsymbol{1}$ (see for example \cite{Erler:2019loq}). Therefore the consistency of the full quantum open-closed theory is codified by the nilpotency requirement \footnote{\color{Black} Throughout the paper, $[\cdot,\cdot]$ represents the graded commutator.}
 \begin{align}
({\bfn}+\bfU)^2=\frac12[{\bfn}+\bfU,{\bfn}+\bfU]=0\quad\rightarrow\quad  \frac12\left(S_{\rm oc},S_{\rm oc}\right)+\Delta S_{\rm oc}=0.\label{nilp}
 \end{align}
 \\
  This is the main result of this note which is organized as follows. In section \ref{sec:2} we review the standard formulation of open-closed SFT by Zwiebach and we define the open and closed multistring products.  In section \ref{sec:3} we formulate these multistring products as (generalized) co-derivations acting on the open-closed tensor algebra and its corresponding group-like element. Then we show how to write the action as a sum of two WZW-like contributions. Finally we package everything into single open-closed structures thus providing the most compact formulation. In section \ref{sec:4} we show how the BV quantum master equation is solved provided \ch{that} the open-closed co-derivation $\bfn$ solves \eqref{nilp} and we examine it order by order in the string coupling constant $\kappa$, retrieving the expected consistency conditions for the sewing of Riemann surfaces which ensure the decoupling of BRST exact states. 
  We conclude in \ref{sec:5} by discussing the advantages of this new formulation and propose some direction for future research.
   
    \section{Open-closed string field theory }\label{sec:2}
    
    Open-closed SFT has two independent dynamical variables: the closed and the open string field, which are upgraded to fields and antifields according to the Batalin-Vilkovisky (BV) quantization.
    The BV closed string is understood as $\Phi=\phi^a c_a$, where $\phi^a$ are the target space BV fields and antifields and $c_a$ are a basis for the closed string Hilbert space associated to the starting closed string background CFT$_0$, subject to level matching
    \begin{align}
    (b_0-\bar b_0)\,c_a=(L_0-\bar L_0)\, c_a=0.
    \end{align}
    In this space there is an odd symplectic form defined by
    \begin{align}
    \wc(\Phi_1,\Phi_2)\coloneqq (-1)^{d(\Phi_1)}\langle\Phi_1,c_0^-\Phi_2\rangle_{\rm c},
    \end{align}
    where $\langle \cdot, \cdot\rangle_{\rm c}$ is the bulk BPZ inner product and where $d(\Phi)$ is the sum of the BV degree of $\phi^a$ and the degree of the basis element $c_a$ which is defined to be its ghost number minus 2.  This is assigned such that the total BV string field has degree zero. The classical closed string field is spanned by basis vectors of degree zero. The symplectic form is graded-antisymmetric 
    \begin{align}
    \wc(\Phi_1,\Phi_2)=-(-1)^{d(\Phi_1)\,d(\Phi_2)}\wc(\Phi_2,\Phi_1).
    \end{align}
    The BV open string field is similarly written as $\Psi=\psi^a o_a$, where $\psi^a$ are the open BV fields and $o_a$ is a basis of the boundary fields of the open string Hilbert space associated to the initial D-brane system BCFT$_0$. The open string symplectic form is defined as
    \begin{align}
    \wo(\Psi_1,\Psi_2)\coloneqq (-1)^{d(\Psi_1)}\langle\Psi_1,\Psi_2\rangle_{\rm o},
    \end{align}
    where $\langle \cdot, \cdot\rangle_{\rm o}$ is the boundary BPZ inner product \ch{(possibly including a trace over Chan-Paton factors)} and $d(\Psi)$ is the sum of the BV degree of $\psi^a$ and the degree of the basis element $o_a$ which is defined to be its ghost number minus 1.  This is assigned such that the total BV string field has degree zero and the classical open string field is spanned by basis vectors of degree zero. 
    
    The BV quantum $(\hbar=1)$ master action is given by
    \begin{equation}
        S_{\rm oc}[\Phi,\Psi]=\sum_{g=0}^{\infty}\sum_{b=0}^{\infty}\kappa^{2g+b-2}\sum_{k=0}^{\infty}\sum_{\{p_1,...,p_b\}=0}^{\infty}\frac{1}{b!k!(p_1)\cdots(p_b)}\mathcal{A}^{g,b}_{k;\{p_1,...,p_b\}}\left(\Phi^{\wedge k}\otimes'\Psi^{\odot p_1}\wedge'...\wedge' \Psi^{\odot p_b}\right),\label{OC-action}   
    \end{equation}
    where $(p)\coloneqq p+\delta_{p,0}$.
    Without too much specification the vertices $\mathcal{A}^{g,b}_{k;\{p_1,...,p_b\}}$ are off-shell amplitudes associated with  Riemann surfaces of genus $g$ and $b$ boundaries with $k$ bulk punctures and $p_i$ boundary punctures on the $i$-th boundary. Notice that there can be boundaries with no \ch{open-string} insertions.  Every vertex is weighted by a corresponding power of the string coupling constant $\kappa$ according to the worldsheet topological expansion. Crucially, the implicit moduli space integration in the off-shell amplitudes $\mathcal{A}^{g,b}_{k;\{p_1,...,p_b\}}$ is cut-off towards closed and open string degeneration but the details of how this is done are not important for the algebraic structure of the theory. The  symmetrized tensor products $\wedge$  symmetrize over bulk punctures 
    \begin{align}
    \Phi_1\wedge\cdots\wedge\Phi_n \coloneqq\sum_{\sigma\in {\mathbf{S}_n}}(-1)^{\epsilon_\sigma}\Phi_{\sigma(1)}\otimes\cdots\otimes\Phi_{\sigma(n)},
    \end{align}
    where $\mathbf{S}_n$ is the group of permutations. {\color{Black}The direct sums of such multi-string states are elements of the symmetric tensor algebra $\mathcal{SH}_{\rm c}$, defined as
    \begin{equation}
        \mathcal{SH}_{\rm c}\coloneqq \bigoplus_{n\geq0} {\cal H}_{\rm c}^{\wedge n}.
    \end{equation}}
    On every boundary open strings are inserted according to the cyclic tensor product
    \begin{align}
    \Psi_1\odot\cdots\odot\Psi_n \coloneqq\sum_{\sigma\in {\mathbf{Z}_n}}(-1)^{\epsilon_\sigma}\Psi_{\sigma(1)}\otimes\cdots\otimes\Psi_{\sigma(n)},
    \end{align}
    where $\mathbf{Z}_n$ is the group of cyclic permutations. The off-shell amplitudes defined by $\mathcal{A}^{g,b}_{k;\{p_1,...,p_b\}}$ are therefore color-ordered. {\color{Black}We can define cyclic tensor algebra  $\mathcal{CH}_{\rm o}$ as follows
    \begin{equation}
        \mathcal{CH}_{\rm o}\coloneqq \bigoplus_{n\geq0} {\cal H}_{\rm o}^{\odot n}.\label{cyclic-algebra}
    \end{equation}
  This cyclic ordering is present on every boundary and all the boundaries are finally symmetrized by  another symmetrized tensor product $\wedge'$. The extra tensor product $\otimes'$ separates closed and open strings. }\\
    Given the closed and open symplectic forms $\wc$ and $\wo$ it is useful to write the $\mathcal{A}^{g,b}_{k;\{p_1,...,p_b\}}$ using multistring products which are defined as follows. First of all let us consider amplitudes with no open strings but not restricted by the genus or by the number of boundaries. These amplitudes can be expressed as
    \begin{equation}
    \mathcal{A}^{g,b}_{k+1}(\Phi_1\wedge\cdots\wedge\Phi_{k+1})\coloneqq\wc\left(\Phi_1,l^{(g,b)}_{k,0}\left(\Phi_2\wedge\cdots\wedge \Phi_{k+1}\right)\right),
    \end{equation}
    where we have defined the degree odd multi string product 
    \begin{align}
    l^{(g,b)}_{k,0}:\,{\cal H}_{\rm c}^{\wedge k}\,\rightarrow\,{\cal H}_{\rm c},
    \end{align}
    which is cyclic with respect to $\wc$
    \begin{align}
    \wc\left(\Phi_1,l^{(g,b)}_{k,0}\left(\Phi_2\wedge\cdots\wedge \Phi_{k+1}\right)\right)=(-1)^\epsilon \wc\left(\Phi_{k+1},l^{(g,b)}_{k,0}\left(\Phi_1\wedge\cdots\wedge \Phi_{k}\right)\right)
    \end{align}
    and is BPZ odd
    \begin{align}
    \wc\left(\Phi_1,l^{(g,b)}_{k,0}\left(\Phi_2\wedge\cdots\wedge \Phi_{k+1}\right)\right)=-(-1)^{d(\Phi_1)} {\color{black}\wc}\left(l^{(g,b)}_{k,0}\left(\Phi_1\wedge\cdots\wedge \Phi_{k}\right),\Phi_{k+1}\right).\label{bpz-cycl-closed}
    \end{align}
    All the remaining amplitudes have at least one open string in at least one boundary and therefore they can be  written using the open string symplectic form, by picking \ch{an arbitrary} boundary with at least one open string puncture. \ch{This}, thanks to full boundary symmetrization, can always be \ch{conventionally} taken to \ch{correspond to} the last in the arguments of $\mathcal{A}^{g,b}$, that is \ch{the $b$-th argument} with $p_b\geq1$
    \begin{align}
    &\mathcal{A}^{g,b}_{k;\{p_1,...,p_b\}}\left(\Phi_1\wedge\cdots\wedge\Phi_k\otimes'\Psi_{1,1}\odot\cdots\odot\Psi_{1, p_1}\,\wedge'...\wedge' \Psi_{b,1}\odot\cdots\odot\Psi_{b, p_b}\right)\nonumber\\
    \coloneqq&\wo\left(\Psi_{b,1}, m^{(g,b)}_{k\left[p_1,...,p_{b-1}\right]p_b-1}\left((\Phi)_k\otimes' [\Psi]_{ p_1}\wedge' \cdots\wedge'  [\Psi]_{ p_{b-1}}\otimes''\Psi_{b,2}\otimes\cdots\otimes\Psi_{b,p_b}\right)\right),
    \end{align}
    where to lighten the notation we have written 
    \begin{align}
    (\Phi)_k&\coloneqq\Phi_1\wedge\cdots\wedge\Phi_k\0\\
    [\Psi]_{p_j}&\coloneqq\Psi_{j,1}\odot\cdots\odot\Psi_{j, p_j}
    \end{align}
    and the normalizations due to cyclicity are defined by $(p)\coloneqq p+\delta_{p,0}$. {\color{Black}Notice that we have introduced the $\otimes''$ to separate the special boundary from the others.} The multistring products with open string output $m^{(g,b)}$ are odd linear maps
    \begin{align}\label{mdef}
    m^{(g,b)}_{k\left[p_1,...,p_{b-1}\right]p_b}: \,{\cal H}_{\rm c}^{\wedge k}\otimes'{\cal H}_{\rm o}^{\odot p_1}\wedge'\cdots\wedge'{\cal H}_{\rm o}^{\odot p_{b-1}}\otimes''{\cal H}_{\rm o}^{\otimes p_b}\rightarrow\,{\cal H}_{\rm o},
    \end{align}
     which are cyclic  
    \begin{align}
    &\wo\left(\Psi_{b,1}, m^{(g,b)}_{k\left[p_1,...,p_{b-1}\right]p_b-1}\left((\Phi)_k\otimes' [\Psi]_{ p_1}\wedge' \cdots\wedge'  [\Psi]_{ p_{b-1}}\otimes'' \Psi_{b,2}\otimes\cdots\otimes\Psi_{b,p_b}\right)\right)\\
    =&(-1)^\epsilon\wo\left(\Psi_{b,p_b}, m^{(g,b)}_{k\left[p_1,...,p_{b-1}\right]p_b-1}\left((\Phi)_k\otimes' [\Psi]_{ p_1}\wedge' \cdots\wedge'  [\Psi]_{ p_{b-1}}\otimes'' \Psi_{b,1}\otimes\cdots\otimes\Psi_{b,p_b-1}\right)\right)\0
    \end{align}
    and BPZ odd 
    \begin{align}
    &\wo\left(\Psi_{b,1}, m^{(g,b)}_{k\left[p_1,...,p_{b-1}\right]p_b-1}\left((\Phi)_k\otimes' [\Psi]_{ p_1}\wedge' \cdots\wedge'  [\Psi]_{ p_{b-1}}\otimes'' \Psi_{b,2}\otimes\cdots\otimes\Psi_{b,p_b}\right)\right)\label{open-cycl1}\\
    =&-(-1)^\epsilon\wo\left( m^{(g,b)}_{k\left[p_1,...,p_{b-1}\right]p_b-1}\left((\Phi)_k\otimes' [\Psi]_{ p_1}\wedge' \cdots\wedge'  [\Psi]_{ p_{b-1}}\otimes'' \Psi_{b,1}\otimes\cdots\otimes\Psi_{b,p_b-1}\right),\Psi_{b,p_b}\right)\0.
    \end{align}
    Obviously there is nothing special about the choice of \ch{the} special boundary $b$ and this has the consequence that the $m^{(g,b)}$ products have extra properties obtained by expressing the same amplitude with different choice of special boundary
    \begin{align}
    &\wo\left(\Psi_{b,1}, m^{(g,b)}_{k\left[p_1,...,p_{b-1}\right]p_b-1}\left((\Phi)_k\otimes' [\Psi]_{ p_1}\wedge' \cdots\wedge'  [\Psi]_{ p_{b-1}}\otimes'' \Psi_{b,2}\otimes\cdots\otimes\Psi_{b,p_b}\right)\right)\0\\
    =&(-1)^\epsilon\,\wo\left(\Psi_{1,1}, m^{(g,b)}_{k\left[p_2,...,p_b\right]p_1-1}\left((\Phi)_k\otimes' [\Psi]_{ p_2}\wedge' \cdots\wedge'  [\Psi]_{ p_b}\otimes''\Psi_{1,2}\otimes\cdots\otimes\Psi_{1,p_1}\right)\right)\label{open-cycl2}\\
    =&\mathcal{A}^{g,b}_{k;\{p_1,...,p_b\}}\left((\Phi)_k\otimes'[\Psi]_{p_1}\wedge'...\wedge' [\Psi]_{p_b}\right)\0.
    \end{align}
    Finally, it is useful to define dual products $l^{(g,b)}$ with a closed string output such that the same open-closed amplitude can be computed with the open or closed symplectic form 
    \begin{align}
    &\wo\left(\Psi_{b,1}, m^{(g,b)}_{k\left[p_1,...,p_{b-1}\right]p_b-1}\left((\Phi)_k\otimes' [\Psi]_{ p_1}\wedge' \cdots\wedge'  [\Psi]_{ p_{b-1}}\otimes'' \Psi_{b,2}\otimes\cdots\otimes\Psi_{b,p_b}\right)\right)\0\\
    \coloneqq&(-1)^\epsilon \wc\left(\Phi_1,l^{(g,b)}_{k-1,[p_1,\cdots, p_b]}\left(\Phi_2\wedge\cdots\wedge\Phi_k\otimes'[\Psi]_{p_1}\wedge'\cdots\wedge'[\Psi]_{p_b}\right)\right)\label{dual}\\
    =&\mathcal{A}^{g,b}_{k;\{p_1,...,p_b\}}\left((\Phi)_k\otimes'[\Psi]_{p_1}\wedge'...\wedge' [\Psi]_{p_b}\right)\0.
    \end{align}
    With the definitions and properties given above, the open-closed BV action \eqref{OC-action} can be written as
    \begin{align}
    &S_{\rm oc}[\Phi,\Psi]=\sum_{g,b=0}^{\infty}\kappa^{2g+b-2}\sum_{k=0}^{\infty}\frac{1}{b!}{\ch{\Bigg [}}\frac1{(k+1)!}
    \wc\left(\Phi,l^{(g,b)}_{k,0}\left(\Phi^{\wedge k}\right)\right)\label{OCact}\\
    &+\ch{\sum_{\substack{\{p_1,...,p_{b-1}\}=0\\ {p_b=1}}}^{\infty}}\frac{\ch{C({p_1,\ldots, p_b})}}{k!\,(p_1)\cdots(p_b)}
    \wo\left(\Psi,m^{(g,b)}_{k\left[p_1,...,p_{b-1}\right]p_b-1}\left(\Phi^{\wedge k}\otimes'\Psi^{\odot p_1}\cdots\Psi^{\odot p_{b-1}}\otimes''\Psi^{\otimes (p_b-1)}\right)\right){\ch{\Bigg ]}}.\0
    \end{align}
	\ch{Note that since we were forced to restrict the sum over $p_b$ to start from 1 (so that the special boundary contains always at least one insertion), the summation would no longer automatically produce the correct symmetry factor in front of couplings containing at least one empty boundary. This needs to be corrected by including into the action the factor $C({p_1,\ldots, p_b})$ which we define as
	\begin{align}
	C({p_1,\ldots, p_b}) = \frac{\text{$\#$ of inequiv.\ rearrangements of the $b$-tuple $(p_1,\ldots,p_b)$ }}{\text{$\#$ of inequiv.\ rearrangements of the $b$-tuple $(p_1,\ldots,p_b)$ s.t.\ the last entry $\neq 0$}}\,.
	\end{align}
	Using elementary combinatorics, one can show that $C({p_1,\ldots, p_b})$ can in fact be expressed as
	\begin{align}
	C({p_1,\ldots, p_b})= \frac{b}{b-b_0}\,,\label{eq:Cfactor}
	\end{align}
	where $0\leq b_0<b$ denotes the number of boundaries with zero open-string field insertions. We will observe below that the presence of this combinatorial factor will be automatically accounted for when working in the WZW co-derivation formalism.
}
    Notice that, using the open-closed relation \eqref{dual} it is possible to write the action in different equivalent ways. \ch{For example,} we can isolate the terms with only open strings
   \ch{ \begin{align}
    &S_{\rm oc}[\Phi,\Psi]=\sum_{g,b=0}^{\infty}\kappa^{2g+b-2}\frac{1}{b!}{\Bigg [}\sum_{\{p_1,...,p_b\}=0}^{\infty}\frac{1}{(p_1)\cdots(p_b)}\sum_{k=0}^{\infty}\frac1{(k+1)!}\0\\
    &\hspace{7cm}  \wc\left(\Phi,l^{(g,b)}_{k,[p_1,\cdots ,p_b]}\left(\Phi^{\wedge k}\otimes \Psi^{\odot p_1}\cdots\Psi^{\odot p_b}\right)\right)+\\
    &\hspace{1cm}+\sum_{\substack{\{p_1,...,p_{b-1}\}=0\\ p_b=1}}^{\infty}\frac{C({p_1,\ldots, p_b})}{(p_1)\cdots(p_b)}\wo\left(\Psi,m^{(g,b)}_{0\left[p_1,...,p_{b-1}\right]p_b-1}\left(\Psi^{\odot p_1}\cdots\Psi^{\odot p_{b-1}}\otimes''\Psi^{\otimes (p_b-1)}\right)\right){\Bigg ]}\0\,.
    \end{align}}
Other rewritings (for example choosing differently the special boundary) are also possible. {\color{Black}It is one of the aim of this paper to give a  presentation of the OC action which is invariant under all of the above-mentioned rewritings, based on an appropriately defined symplectic form.}

 \section{Co-algebraic open-closed structure}\label{sec:3}

    To start with we define the open-closed tensor algebra as
    \begin{align}
    \mathcal{SH}_{\rm c}\otimes' \mathcal{SCH}_{\rm o}\coloneqq \left[\bigoplus_{k\geq0} {\cal H}_{\rm c}^{\wedge k}\right]\otimes'\left[\bigoplus_{b\geq0}\left(\bigoplus_{p_1\geq0}{\cal H}_{\rm o}^{\odot p_1}\right)\wedge'\cdots\wedge'\left(\bigoplus_{p_b\geq0}{\cal H}_{\rm o}^{\odot p_b}\right)\right],
    \end{align}
    {\color{Black} this formula gives a definition of $\mathcal{SCH}_{\rm o}$ as the direct sum of the symmetric tensor product of cyclic tensor algebras, defined in \eqref{cyclic-algebra}.}
    To manipulate the action it is useful to introduce the open-closed group-like element 
    \begin{align}
    {\cal G}\coloneqq e^{\wedge \Phi}\otimes'\,e^{\wedge' {\cal C}(\Psi)},
    \end{align}
    where we have defined the cyclic group-like element\footnote{The symmetrized and cyclicized group like elements $e^{\wedge \chi}$ and ${\cal C}(\chi)$ are the same when expressed in terms of the simple tensor product $\otimes$, $e^{\wedge \chi}={\cal C}(\chi)=\frac1{1-\otimes\chi}=\sum_{k\geq0}\chi^{\otimes k}$.}
    \begin{align}
    {\cal C}(\Psi)\coloneqq\sum_{l=0}^\infty  \frac1{(l)}\Psi^{\odot l}=1+\log(1-\odot\Psi).
    \end{align}
    Explicitly we can write
    \begin{align}
    {\cal G}=\left[\sum_{k\geq0}\frac1{k!} \Phi^{\wedge k}\right]\otimes'\left[\sum_{b\geq0}\frac1{b!}\left(\sum_{p_1\geq0}\frac{1}{(p_1)}\Psi^{\odot p_1}\right)\wedge'\cdots\wedge'\left(\sum_{p_b\geq0}\frac1{(p_b)}\Psi^{\odot p_b}\right)\right].
    \end{align}
    As we are now going to see ${\cal G}$ is the correct reservoir for symmetrized closed strings, together with cyclicized open strings taken from symmetrized boundaries, that appear in the action. 
    
    The multistring products $l^{(g,b)}$ and $m^{(g,b)}$ can be upgraded to {\color{Black}odd} \ch{coderivation-like maps} on $\mathcal{SH}_{\rm c}\otimes'\mathcal{SCH}_{\rm o}$. This is straightforward for the $l^{(g,b)}$ products because their inputs are already in $\mathcal{SH}_{\rm c}\otimes' \mathcal{SCH}_{\rm o}$ and we have
    \begin{align}
    \boldsymbol{l}_{k[p_1,\cdots, p_b]}^{(g,b)}:\mathcal{SH}_{\rm c}\otimes' \mathcal{SCH}_{\rm o}\rightarrow\mathcal{SH}_{\rm c}\otimes' \mathcal{SCH}_{\rm o},
    \end{align}
    where the \ch{defining} action on the group-like element ${\cal G}$ is as follows
    \begin{align}
    \boldsymbol{l}_{k[p_1,\cdots ,p_b]}^{(g,b)}{\cal G}=&\sum_{k'\geq k}^{\infty}\sum_{b'\geq b}^{\infty}\sum_{\{p'_{b+1},...,p'_{b'}\}\geq0}\dfrac{1}{b'!k'!\left((p_1)\cdots(p_b)(p'_{b+1})\cdot \cdot\cdot (p'_{b'})\right)}\binom{k'}{k}\binom{b'}{b}\\
    &\left(l_{k[p_1,\cdots, p_b]}^{(g,b)}\left(\Phi^{\wedge k}\otimes' \Psi^{\odot p_1}\wedge'\cdots\wedge'\Psi^{\odot p_b}\right)\wedge \Phi^{\wedge (k'-k)} \otimes'  \Psi^{\odot p'_{b+1}}\wedge'\cdots\wedge' \Psi^{\odot p'_{b'}}\right),
    \end{align}
    where the binomial coefficients account for the choice of $k$ bulk punctures out of $k'$ and $b$ boundaries out of $b'$.     Notice in particular that it behaves as a co-derivation for the closed string inputs and for the open strings on the boundaries which are \ch{not acted upon}. On the other hand the number of open string insertions on the boundaries that enter into the multistring product has to precisely match the indices $\{p_1,\cdots,p_b\}$. The reader can easily extend this definition to a generic element of $\mathcal{SH}_{\rm c}\otimes' \mathcal{SCH}_{\rm o}$.

    We can then sum on all possible inputs and, at fixed genus and boundaries, define
    \begin{align}
     \boldsymbol{l}^{(g,b)}\coloneqq\sum_{k,\left\{p_1,...,p_b\right\}}\boldsymbol{l}^{(g,b)}_{k\left[p_1,...,p_b\right]}.
    \end{align}
    This is an example of a{\color{Black}n odd} cyclic closed string co-derivation which obeys
    \begin{align}
    \wc\left(\pi_{10}\, \boldsymbol{l}^{(g,b)}\,\boldsymbol{a}_{\rm c}\,{\cal G}\,,\,\pi_{10}\boldsymbol{b}_{\rm c}\,{\cal G}\right)=-(-1)^{d(a)}\wc\left(\pi_{10} \,\boldsymbol{a}_{\rm c}\,{\cal G}\,,\,\pi_{10}\boldsymbol{l}^{(g,b)}\boldsymbol{b}_{\rm c}\,{\cal G}\right),\label{clcycl}
    \end{align}
where $\pi_{10}$ projects \ch{onto} a single copy of ${\cal H}_{\rm c}$ inside $\mathcal{SH}_{\rm c}\otimes' \mathcal{SCH}_{\rm o}$ and $(\boldsymbol{a}_{\rm c},\boldsymbol{b}_{\rm c})$ are any two closed string co-derivations ({\it i.e.} with closed string output). This is a consequence of \ch{a relation analogous to \eqref{bpz-cycl-closed} which holds for the products $l^{(g,b)}$}.

    We \ch{would now} like to upgrade also the products $m^{(g,b)}$ (with open string output) to  $\mathcal{SH}_{\rm c}\otimes' \mathcal{SCH}_{\rm o}$. An immediate difficulty for doing so is that the \ch{products} $m^{(g,b)}_{k[p_1,\cdots, p_{b-1}]p_b}$ are defined with non-cyclic inputs on the special boundary $b$, \eqref{mdef}. However it is easy to realize that co-derivations on the tensor algebra are automatically co-derivations on the cyclic tensor algebra. In particular if $\boldsymbol{c}_i$ is a co-derivation on ${\cal TH}$ it will also act as a \ch{coderivation-like map} on ${\cal CH}$: \ch{for $i\leq j$ we have}
    \begin{equation}
    \begin{split}
    \boldsymbol{c}_{i}\left(\Psi_{1}\odot\cdot\cdot\cdot \odot \Psi_{j}\right)&= \boldsymbol{c}_{i}\left(\sum_{\sigma\in \mathbf{Z}_{j}}\Psi_{\sigma(1)}\otimes\cdot\cdot\cdot \otimes \Psi_{\sigma(j)}\right)=\\
    &=\sum_{\sigma\in \mathbf{Z}_{j}}c_{i}\left(\Psi_{\sigma(1)}\otimes\cdot\cdot\cdot \otimes \Psi_{\sigma(i)}\right)\otimes \Psi_{\sigma(i+1)}\otimes \cdot\cdot\cdot \otimes\Psi_{\sigma(l)}+\\
    &\qquad \vdots \\
    &+\sum_{\sigma\in \mathbf{Z}_{j}}\Psi_{\sigma(1)}\otimes \cdot\cdot\cdot \otimes\Psi_{\sigma(j-i)}\otimes c_{i}\left(\Psi_{\sigma(j-i+1)}\otimes\cdot\cdot\cdot \otimes \Psi_{\sigma(j)}\right)=\\
    &=c_{i}\left(\Psi_{1}\otimes\cdots \otimes \Psi_{i}\right)\odot \Psi_{i+1}\odot\cdots\odot \Psi_{j}\\
    &+c_{i}\left(\Psi_{2}\otimes\cdots \otimes \Psi_{i+1}\right)\odot \Psi_{i+2}\odot\cdots \Psi_{j}\odot \Psi_{1}\\
    &\qquad \vdots\\
    &+c_{i}\left(\Psi_{j}\otimes\Psi_1\otimes\cdots \otimes \Psi_{i-1}\right)\odot \Psi_{i}\odot\cdots\odot \Psi_{j-1},\\
    \end{split}
    \end{equation}
    where we have assumed degree even entries (otherwise some obvious signs appear).
   Therefore, \ch{one can consistently restrict}
    \begin{align}
    \boldsymbol{c}_i:&\,{\cal TH}\to{\cal TH}\0\\
    &\hspace{0.5cm}\ch{\downarrow}\\
    \boldsymbol{c}_i:&\,{\cal CH}\to{\cal CH}\0.
    \end{align}
    Notice in particular that \ch{for $i\leq j$} we have
    \begin{equation}
    \label{eqn: codcyc}
    \boldsymbol{c}_{i}\left(\Psi^{\odot j}\right)=(j)\,c_{i}\left(\Psi^{\otimes i}\right)\odot \Psi^{\odot j-i}.
    \end{equation}
    This suggests to define the {\color{Black}odd} co-derivation with open string output
    \begin{align}
    \boldsymbol{m}^{(g,b)}_{k[p_1,\cdots, p_{b-1}]p_b}: \mathcal{SH}_{\rm c}\otimes' \mathcal{SCH}_{\rm o}\to\mathcal{SH}_{\rm c}\otimes' \mathcal{SCH}_{\rm o},
    \end{align}
    in such a way that its action on the open-closed group-like element will be
    \begin{equation}
    \label{eqn: defm}
    \begin{split}
    &\boldsymbol{m}^{(g,b)}_{k\left[p_1,...,p_{b-1}\right]p_b}{\cal G}=\sum_{k'\geq k}\sum_{b'\geq b}\sum_{p'_b\geq p_b}\ch{\sum_{\{p'_{b+1},...,p'_{b'}\}\geq0}}\dfrac{1}{b'!k'!}\,\frac{1}{(p_1)\cdot \cdot\cdot (p_{b-1})} \, \frac1{(p'_b)}\,\frac1{(p'_{b+1})\cdots (p'_{b'})}\ch{\times}\\
    &\quad\quad\quad \quad\quad\quad\quad \ch{\times}\binom{k'}{k}\,\binom{b'}{b}\,b\, (p'_b)\ch{\times} \\
    & \ch{\times}\left(\Phi^{\wedge k'-k}\otimes' m^{(g,b)}_{k\left[p_1,...,p_{b-1}\right]p_b}\left(\Phi^{\wedge k}\otimes' \Psi^{\odot p_1}\wedge'...\wedge'\Psi^{\odot p_{b-1}}\otimes''\Psi^{\otimes p_b}\right) 
    \odot \Psi^{\odot p'_b-p_b}
     \wedge' \Psi^{\odot p'_{b+1}}\wedge'...\wedge' \Psi^{\odot p'_{b'}}\right),
    \end{split}
    \end{equation}
    where, to help the visualization, the first line contains the coefficients of the open-closed group-like element, the second line contains the combinatorial factors associated to choosing $k$ closed strings out of $k'$, $b$ boundaries out of $b'$ and choosing one special boundary out of $b$. Finally the coefficient $(p'_b)$ is the result of \eqref{eqn: codcyc}.  Notice the cancellation $\frac{1}{(p'_b)}\,(p'_b)=1$.
    The explicit action on a generic element of $\mathcal{SH}_{\rm c}\otimes' \mathcal{SCH}_{\rm o}$ is straightforwardly obtained given the understood symmetrizations and cyclicizations. It is however rather cloggy and not particularly instructive to write it down.
    
Analogously to what we have done with the closed string products $\bfl^{(g,b)}$, we can now sum over the full open and closed insertions at fixed genus and boundaries
\begin{align}
\bfm^{(g,b)}\coloneqq\sum_{k\geq0}\sum_{p_1\cdots p_b\geq0}\boldsymbol{m}^{(g,b)}_{k\left[p_1,...,p_{b-1}\right]p_b}.
\end{align} 
 The \ch{cyclicity} properties further imply the relations
    \begin{align}
    \wo\left(\pi_{01} \boldsymbol{m}^{(g,b)}\boldsymbol{a}_{\rm o}\,{\cal G},\pi_{01}\boldsymbol{b}_{\rm o}\,{\cal G}\right)=-(-1)^{d(a)}\wo\left(\pi_{01} \boldsymbol{a}_{\rm o}\,{\cal G},\pi_{01}\boldsymbol{m}^{(g,b)}\boldsymbol{b}_{\rm o}\,{\cal G}\right), \label{opcycl}
    \end{align}
for any pair of open string co-derivations $(\boldsymbol{a}_{\rm o},\boldsymbol{b}_{\rm o})$. This is a consquence of \eqref{open-cycl1} and \eqref{open-cycl2}.
    
By construction the obtained open and closed string co-derivations $\bfm^{(g,b)}$ and $\bfl^{(g,b)}$ are not independent but obey the open-closed \ch{duality} relation 
\begin{align}
\wc\left(\pi_{10}\,\bfl^{(g,b)}\,\boldsymbol{a}_{\rm o}\,{\cal G}\,,\, \pi_{10}\,\boldsymbol{b}_{\rm c}\,{\cal G}\right)&=-(-1)^{d(a)}\wo\left(\pi_{01}\,\boldsymbol{a}_{\rm o}\,{\cal G}\,,\, \pi_{01}\,\bfm^{(g,b)}\,\boldsymbol{b}_{\rm c}\,{\cal G}\right) \label{opclcycl1} \\
\wo\left(\pi_{01}\,\bfm^{(g,b)}\,\boldsymbol{a}_{\rm c}\,{\cal G}\,,\, \pi_{01}\,\boldsymbol{b}_{\rm o}\,{\cal G}\right)&=-(-1)^{d(a)}\wc\left(\pi_{10}\,\boldsymbol{a}_{\rm c}\,{\cal G}\,,\, \pi_{10}\,\bfl^{(g,b)}\,\boldsymbol{b}_{\rm o}\,{\cal G}\right),\label{opclcycl2}
\end{align}
where $(\boldsymbol{a}_{\rm o,c}, \boldsymbol{b}_{\rm o, c})$ are open or closed co-derivations. This is a consequence of \eqref{dual}.

\subsection{BV Action in WZW form}

To write the action in a compact form we use a standard trick. We first define \ch{an interpolation for both} the open and closed string fields: \ch{namely} $\Phi(t)$ \ch{such that} $\Phi(0)=0$, $\Phi(1)=\Phi$ and $\Psi(t)$ \ch{such that} $\Psi(0)=0$, $\Psi(1)=1$. Then we write the action \eqref{OCact} as the integral of a total derivative by inserting $\int_0^1dt\partial_t$ in front.
\ch{Let us} start with the simpler term containing only closed strings
\begin{align}
 &\sum_{g,b=0}^{\infty}\kappa^{2g+b-2}\sum_{k=0}^{\infty}\frac{1}{(k+1)!b!}\left(\int_0^1dt\partial_t\right)
    \wc\left(\Phi,l^{(g,b)}_{k,0}\left(\Phi^{\wedge k}\right)\right)\0\\
    =&\sum_{g,b=0}^{\infty}\kappa^{2g+b-2}\sum_{k=0}^{\infty}\int_0^1dt\frac{k+1}{(k+1)!b!} \wc\left(\dot\Phi,l^{(g,b)}_{k,0}\left(\Phi^{\wedge k}\right)\right)\0\\
    =&\sum_{g,b=0}^{\infty}\kappa^{2g+b-2}\frac1{b!}\int_0^1\,dt \wc\left(\dot\Phi,\left[\sum_{k=0}^{\infty}\frac{1}{k!}l^{(g,b)}_{k,0}\left(\Phi^{\wedge k}\right)\right]\right)\ch{\,.}\label{WZWc1}
\end{align}
\ch{We} keep it as it is for the time being and we consider the other term in \eqref{OCact} containing  open strings (excluding for \ch{now} the sum over topologies)
\begin{align}
\ch{\sum_{\substack{k,\{p_1,...,p_{b-1}\}=0\\ p_b=1}}^{\infty}}\frac{\ch{C(p_1,\ldots,p_b)}}{b!k!(p_1)\cdots(p_b)}
   \left(\int_0^1dt\partial_t\right) \wo\left(\Psi,m^{(g,b)}_{k\left[p_1,...,p_{b-1}\right]p_b-1}\left(\Phi^{\wedge k}\otimes'\Psi^{\odot p_1}\cdots\Psi^{\odot p_{b-1}}\otimes''\Psi^{\otimes (p_b-1)}\right)\right).\label{actOP}
   \end{align}
   The $t$-derivative now acts separately on $\Phi(t)$ and $\Psi(t)$. We start writing down the contribution from $\Phi(t)$ which is given by
   \begin{align}
\ch{\sum_{\substack{k=1,\{p_1,...,p_{b-1}\}=0\\ p_b=1}}^{\infty}}\frac{k\ch{C(p_1,\ldots,p_b)}}{b!k!(p_1)\cdots(p_b)}
   \int_0^1dt\, \wo\left(\Psi,m^{(g,b)}_{k\left[p_1,...,p_{b-1}\right]p_b-1}\left(\dot\Phi\wedge\Phi^{\wedge k-1}\otimes'\Psi^{\odot p_1}\cdots\Psi^{\odot p_{b-1}}\otimes''\Psi^{\otimes (p_b-1)}\right)\right),
   \end{align}
   now we use the open-closed \ch{duality} \eqref{dual} and rewrite it using the closed symplectic form \ch{as}  
     \begin{align}
\sum_{\ch{k=1},\{p_1,...,p_b\}\ch{=0}}\frac{k}{b!k!(p_1)\cdots(p_b)}
   \int_0^1dt\, \wc\left(\dot\Phi,l^{(g,b)}_{k-1\left[p_1,...,p_{b-1},p_b\right]}\left(\Phi^{\wedge k-1}\otimes'\Psi^{\odot p_1}\cdots\Psi^{\odot p_b}\right)\right)\,.\label{WZWc2}
   \end{align} 
\ch{Notice that the combinatorial factor $C(p_1,\ldots,p_b)$ disappeared from the last expression because now all the summation indices $p_1,\ldots,p_b$ are allowed to be zero.} We then realize that we can combine this term (adding the understood sum over topologies $\sum_{g,b}\kappa^{2g+b-2}$) with \eqref{WZWc1} to obtain
  \begin{align}
\eqref{WZWc1}+\sum_{g,b}\kappa^{2g+b-2}\eqref{WZWc2}&= \int_0^1dt\,\sum_{g,b}\kappa^{2g+b} \,\frac{\wc}{\kappa^2}\left(\pi_{10}\boldsymbol{\partial}_t\,{\cal G}\,,\,\pi_{10} \bfl^{(g,b)}\,{\cal G}\right)\0\\
&= \int_0^1dt\,\frac{\wc}{\kappa^2}\left(\pi_{10}\boldsymbol{\partial}_t\,{\cal G}\,,\,\pi_{10} \bfl\,{\cal G}\right),
  \end{align}
  where we have defined the total closed string co-derivation
  \begin{align}
\bfl\coloneqq \sum_{g,b}\kappa^{2g+b}\bfl^{(g,b)}.  \label{l-expand}
  \end{align}
After this treatment we remain with the $\partial_t$ acting on $\Psi(t)$ in \eqref{actOP}. \ch{This will of course act only on the $b-b_0$ boundaries which contain at least one open-string field insertion. Using the formula \eqref{eq:Cfactor} for $C(p_1,\ldots,p_b)$, we observe that acting with the derivative $\p_t$ therefore turns the combinatorial factor $C(p_1,\ldots,p_b)$  into a factor of $b$.} Thanks to the cyclicity relations \eqref{open-cycl1} and \eqref{open-cycl2} and \ch{due} to the explicit summations on open string insertions we \ch{therefore} produce an overall factor $b(p_b)$ obtaining in total, cfr \eqref{eqn: defm}
\begin{align}
&\sum_{g,b}\sum_{k,\{p_1,...,p_b,p_b\geq1\}}\frac{\kappa^{2g+b-2}\,b(p_b)}{b!k!(p_1)\cdots(p_b)}
   \int_0^1dt\, \, \wo\left(\dot\Psi,m^{(g,b)}_{k\left[p_1,...,p_{b-1}\right]p_b-1}\left(\Phi^{\wedge k}\otimes'\Psi^{\odot p_1}\cdots\Psi^{\odot p_{b-1}}\otimes''\Psi^{\otimes (p_b-1)}\right)\right)\0\\
 =&  \sum_{g,b}\kappa^{2g+b-1}\int_0^1 dt\,  \frac\wo\kappa\left(\pi_{01}\,\boldsymbol{\partial}_t\,{\cal G}\,,\,\pi_{01}\bfm^{(g,b)}\,{\cal G}\right)\\
 =&\int_0^1 dt\,  \frac\wo\kappa\left(\pi_{01}\,\boldsymbol{\partial}_t\,{\cal G}\,,\,\pi_{01}\bfm\,{\cal G}\right),
\end{align}
where we have defined the total open string co-derivation
\begin{align}
\bfm\coloneqq\sum_{g,b}\kappa^{2g+b-1} \bfm^{(g,b)}.\label{m-expand}
\end{align}
The full BV master action can thus be written as
\begin{align}
S_{\rm oc}\ch{[}\Phi,\Psi\ch{]}=\int_0^1dt\left(\frac{\wc}{\kappa^2}\left(\pi_{10}\boldsymbol{\partial}_t\,{\cal G}\,,\,\pi_{10} \bfl\,{\cal G}\right)+ \frac\wo\kappa\left(\pi_{01}\,\boldsymbol{\partial}_t\,{\cal G}\,,\,\pi_{01}\bfm\,{\cal G}\right)\right),\label{OC-WZWoc}
\end{align}
where we took the \ch{liberty} of choosing the natural normalizations $\kappa^{-\chi}$ for the closed and open symplectic forms which compute correlators on the sphere $(\chi=2)$ and on the disk $(\chi=1)$ respectively. \\
At last we point out that, as expected from the WZW form, the variation of the action is a pure $t$-boundary term that can be explicitly evaluated as
\begin{align}
   \label{eqn: planarvariation}
        \delta S_{\rm oc}=&\dfrac{\omega_{\rm o}}{\kappa}\left(\pi_{01}\boldsymbol{\delta}_{o}{\cal G},\pi_{01}\boldsymbol{m}\,{\cal G}\right)\Bigr\vert_{0}^{1}+\dfrac{\omega_{\rm c}}{\kappa^{2}}\left(\pi_{10}\boldsymbol{\delta}_{c}\,{\cal G},\pi_{10}\boldsymbol{l}\,{\cal G}\right)\Bigr\vert_{0}^{1}\\
 =&      \dfrac{\omega_{\rm o}}{\kappa}\left(\delta\Psi,\pi_{01}\boldsymbol{m}\,{\cal G}\right)+ \dfrac{\omega_{\rm c}}{\kappa^2}\left(\delta\Phi,\pi_{10}\boldsymbol{l}\,{\cal G}\right) ,
\end{align}
where the generic variation $\delta$ has been naturally upgraded to a co-derivation which variates every entry of the \ch{interpolated group-like element according to the Leibniz rule on the tensor product such that $\delta\Phi(0)=0=\delta\Psi(0)$} and \ch{can be} naturally split into its \ch{open- and closed-string valued} part $\delta\to\boldsymbol{\delta}=\boldsymbol{\delta}_{\rm c}+\boldsymbol{\delta}_{\rm o}$.
\subsection{Open-closed co-derivations and repackaging}\label{sec:2.2}
The form of the action \eqref{OC-WZWoc} is clearly suggesting an even more compact formulation.
First of all we can define a ($\kappa$-dependent) symplectic form in the space ${\cal H}_{\rm c}\oplus{\cal H}_{\rm o}$ as 
\begin{align}
\hw(\Phi_1+\Psi_1\,,\,\Phi_2+\Psi_2)\coloneqq\frac\wc{\kappa^2}(\Phi_1,\Phi_2)+\frac\wo\kappa(\Psi_1,\Psi_2).\label{hatw}
\end{align}
Next we  realize that calling $\chi\coloneqq \Phi+\Psi$ the open-closed string field, this is simply extracted from the open-closed group-like element  $\cal G$ by the projection on a single string $\pi_1\coloneqq \pi_{10}+\pi_{01}$ as
\begin{align}
\pi_1{\cal G}=\chi=\Phi+\Psi.
\end{align}
Then we simply observe that for open and closed string co-derivations $(\bfl, \bfm)$  we obviously have
\begin{align}
\pi_{01}\bfl&=0,\\
\pi_{10}\bfm&=0
\end{align}
so that we can define the open-closed co-derivation
\begin{equation}
\bfn\coloneqq \,\bfl+\bfm
\end{equation}
 to re-write the action \eqref{OC-WZWoc} simply as
 \begin{align}
 S_{\rm oc}\ch{[}\Phi,\Psi\ch{]}=S\ch{[}\chi\ch{]}=\int_0^1dt\,\hw\left(\pi_1\bdel_t\,{\cal G}\,,\,\pi_1\,\bfn\,{\cal G}\right).
 \end{align}
 In the total open-closed Hilbert space ${\cal H}_{\rm c}\oplus{\cal H}_{\rm o}$ it is important to realize that $\bfn$ is a cyclic co-derivation with respect to $\hw$, meaning that given any two open-closed co-derivations $\boldsymbol{a}=\boldsymbol{a}_{\rm c}+\boldsymbol{a}_{\rm o}$ and $\boldsymbol{b}=\boldsymbol{b}_{\rm c}+\boldsymbol{b}_{\rm o}$, as a consequence of (\ref{clcycl}, \ref{opcycl}, \ref{opclcycl1}, \ref{opclcycl2}), we have that
 \begin{align}
 \hw\left(\pi_1\,\bfn\,\boldsymbol{a}\,{\cal G}\,,\,\pi_1\, \boldsymbol{b}\,{\cal G}\right)=-(-1)^{d(a)}\hw\left(\pi_1\,\boldsymbol{a}\,{\cal G}\,,\,\pi_1\, \bfn\,\boldsymbol{b}\,{\cal G}\right).
 \end{align}
 The variation of the action can also be written in compact form by repackaging \eqref{eqn: planarvariation} as
 \begin{align}     \label{OCvary}
 \delta S_{\rm oc}=\hw\left(\pi_1\boldsymbol{\delta}\,{\cal G}\,,\,\pi_1\bfn\,{\cal G}\right)\Bigr\vert_{0}^{1}=\hw\left(\delta\chi\,,\,\pi_1\bfn\,{\cal G}\right).
 \end{align}
 \section{BV quantum master equation and homotopy relations}\label{sec:4}
 Now we would like to establish under which conditions the open-closed SFT action \eqref{OC-action} is consistent, i.e. it is a solution to the  BV master equation ($\hbar=1$)
 \begin{align}
(S,S)+2\Delta S=(S,S)_{\rm c}+ (S,S)_{\rm o}+2\Delta_{\rm c}S+2\Delta_{\rm o}S=0,
 \end{align}
where the open and closed BV structures are given by
\begin{equation}
    \left(\cdot, \cdot\right)_{c}=\kappa^{2}\dfrac{\overleftarrow{\partial}}{\partial\phi^{a}}(\omega_{\rm c})^{ab}\dfrac{\overrightarrow{\partial}}{\partial \phi^{b}},
\end{equation}
\begin{equation}
    \left(\cdot, \cdot\right)_{o}=\kappa \dfrac{\overleftarrow{\partial}}{\partial\psi^{a}}(\omega_{\rm o})^{ab}\dfrac{\overrightarrow{\partial}}{\partial \psi^{b}}.
\end{equation}
and
\begin{equation}
    \Delta_{c}=\dfrac{\kappa^{2}}{2}(-)^{\phi^{a}}(\omega_{\rm c})^{ab}\dfrac{\overrightarrow{\partial}}{\partial \phi^{a}}\dfrac{\overrightarrow{\partial}}{\partial \phi^{b}},
\end{equation}
\begin{equation}
    \Delta_{o}=\dfrac{\kappa}{2}(-)^{\psi^{a}}(\omega_{\rm o})^{ab}\dfrac{\overrightarrow{\partial}}{\partial \psi^{a}}\dfrac{\overrightarrow{\partial}}{\partial \psi^{b}},
\end{equation}
where summation on repeated indices is understood.
The components of the symplectic form with upper indices are defined as 
\begin{align}
(\omega_{\rm c})^{ab}&\coloneqq\omega_{\rm c}(c^a,c^b)\\
(\omega_{\rm o})^{ab}&\coloneqq\omega_{\rm o}(o^a,o^b)
\end{align}
where $c^a$ and $o^a$  are closed and open basis vectors which are dual to $c_a$ and  $o_a$ in the sense that
\begin{align}
\omega_{\rm c} (c^a,c_b)&=-\omega_{\rm c} (c_b,c^a)={\delta^a}_b\\
\omega_{\rm o} (o^a,o_b)&=-\omega_{\rm c} (o_b,o^a)={\delta^a}_b.
\end{align}
Notice also the completeness relations
\begin{align}
\mathbf{1}_{{\cal H}_{\rm c}}&=c_{a}\omega_{\rm c}\left(c^{a}, \cdot\right)=-c^{a}\omega_{\rm c}\left(c_{a},\cdot\right)\label{comp-c}\\
\mathbf{1}_{{\cal H}_{\rm o}}&=o_{a}\omega_{\rm o}\left(o^{a}, \cdot\right)=-o^{a}\omega_{\rm o}\left(o_{a},\cdot\right).\label{comp-o}
\end{align}
We can now define the closed and open Poisson bi-vectors as 
\begin{align}
U_{\rm c}&=\frac{(-1)^{c^a}}{2}c_a\wedge c^a\in {\cal H}_{\rm c}^{\wedge 2}\\
U_{\rm o}&=\frac{(-1)^{o^a}}{2}o_a\wedge o^a\in {\cal H}_{\rm o}^{\wedge 2}.
\end{align} 
They are the inverse of the corresponding symplectic forms in the sense that
\begin{align}
(\wc\otimes \boldsymbol{1})(1\otimes U_{\rm c})&=\boldsymbol{1}_{{\cal H}_{\rm c}}\\
(\wo\otimes \boldsymbol{1})(1\otimes U_{\rm o})&=\boldsymbol{1}_{{\cal H}_{\rm o}},
\end{align}
{\color{Black} which, after applying to a (closed or open) state (of even degree, for simplicity) read
\begin{align}
(\wc\otimes \boldsymbol{1})(\Phi\otimes U_{\rm c})&=\Phi\\
(\wo\otimes \boldsymbol{1})(\Psi\otimes U_{\rm o})&=\Psi,
\end{align}}
as can be easily checked using the completeness relations \eqref{comp-c} and \eqref{comp-o} {\color{Black} (for review see \cite{Erler:2019loq})}.

These objects can be  unified in  open-closed structures  by defining the  open-closed BV string field
\begin{align}
\chi=\chi^a f_a=\Phi^a c_a+\Psi^a o_a.
\end{align}
The total BV bracket is then given by
\begin{align}
(\,\cdot\,,\,\cdot\,)=\dfrac{\overleftarrow{\partial}}{\partial\chi^{a}}\,\hw^{ab}\,\dfrac{\overrightarrow{\partial}}{\partial \chi^{b}}
\end{align}
and the symplectic laplacian
\begin{align}
 \Delta=\frac{1}{2}(-)^{\chi^{a}}\,\hw^{ab}\dfrac{\overrightarrow{\partial}}{\partial \chi^{a}}\dfrac{\overrightarrow{\partial}}{\partial \chi^{b}}.
\end{align}
Notice that in these unified open-closed structures the dependence on the string coupling constant is hidden inside  the open-closed symplectic form $\hw$ \eqref{hatw}. Now the completeness relation is \footnote{Notice that $f^a=(\kappa^2\,c^a, \kappa\,o^a)$.}
\begin{align}\label{OCcompl}
\mathbf{1}_{{\cal H}_{\rm c}\oplus{\cal H}_{\rm o}}&=f_{a}\hw\left(f^{a}, \cdot\right)=-f^{a}\hw\left(f_{a},\cdot\right)
\end{align}
and the open-closed Poisson bivector is given by
\begin{align}
U=\kappa^2\, U_{\rm c}+\kappa\, U_{\rm o}=\frac{(-1)^{f^a}}{2}f_a\wedge f^a,
\end{align}
which is easily shown to obey
\begin{align}
(\hw\otimes \boldsymbol{1})(1\otimes U)&=\boldsymbol{1}_{{\cal H}_{\rm c}\oplus{\cal H}_{\rm o}}.
\end{align}
 \subsection{BV master equation}
 
 Let us see how we can  evaluate the master equation directly at the unified open-closed co-algebra level. Starting with the BV bracket we find
 \begin{align}
 (S_{\rm oc},S_{\rm oc})=&S_{\rm oc}\,\dfrac{\overleftarrow{\partial}}{\partial\chi^{a}}\,\hw^{ab}\,\dfrac{\overrightarrow{\partial}}{\partial \chi^{b}}\,S_{\rm oc}\\
 =&(-1)^{\chi^b}\hw\left(\pi_1 \boldsymbol{\partial}_{\chi^a}\,{\cal G}\,,\,\pi_1 \bfn\,{\cal G}\right)\,\hw^{ab}\,\hw\left(\pi_1 \boldsymbol{\partial}_{\chi^b}\,{\cal G}\,,\,\pi_1 \bfn\,{\cal G}\right)\0\\
 =&(-1)^{\chi^b}\hw\left(f_a\,,\,\pi_1 \bfn\,{\cal G}\right)\,\hw^{ab}\,\hw\left(f_b\,,\,\pi_1 \bfn\,{\cal G}\right)\0\\
 =&-\hw\left(\pi_1 \bfn\,{\cal G}\,,\,\pi_1 \bfn\,{\cal G}\right)\\
 =&-\int_0^1dt\, \partial_t\,\hw\left(\pi_1 \bfn\,{\cal G}(t)\,,\,\pi_1 \bfn\,{\cal G}(t)\right)\0\\
 =&-2\int_0^1dt\,\hw\left(\pi_1 \bfn\boldsymbol{\partial}_t\,{\cal G}\,,\,\pi_1 \bfn\,{\cal G}\right)\0\\
 =&2\int_0^1dt\,\hw\left(\pi_1 \boldsymbol{\partial}_t\,{\cal G}\,,\,\pi_1 \bfn\bfn\,{\cal G}\right)=2\int_0^1dt\,\hw\left(\dot\chi,\,\pi_1 \bfn^2\,{\cal G}\right)
 \end{align}
 where we have used the variation of the action \eqref{OCvary}, the completeness relation \eqref{OCcompl} and we have finally rewritten everything in interpolated way to be able to bring the co-derivation $\bfn$ to the other side.

 Consider now the symplectic laplacian
 \begin{align}
 \Delta S_{\rm oc}=&\frac{1}{2}(-)^{\chi^{a}}\,\hw^{ab}\dfrac{\overrightarrow{\partial}}{\partial \chi^{a}}\dfrac{\overrightarrow{\partial}}{\partial \chi^{b}}S_{\rm oc}\0\\
 =&-\frac{1}{2}\,\hw^{ab}\hw\left(\pi_1 \boldsymbol{\partial}_{\chi^a}\,{\cal G}\,,\,\pi_1 \bfn\boldsymbol{\partial}_{\chi^b}\,{\cal G}\right)\0\\
=&\frac{1}{2}\,\hw\left(f^a\,,\,\pi_1 \bfn\boldsymbol{f}_{a}\,{\cal G}\right),
\end{align}
where we have introduced the $0$-product co-derivation $\boldsymbol{f}_{a}$ which inserts the open-closed basis element $f_a$ in ${\cal G}$ as an open-closed co-derivation.
At this point we can use the consequences of open-closed \ch{duality} (see appendix \ref{app1} for details) to write
\begin{align}
 &\frac{1}{2}\,\hw\left(f^a\,,\,\pi_1 \bfn\boldsymbol{f}_{a}\,{\cal G}\right)\0\\
 =&\frac{1}{2}(-)^{f^{a}}\,\int_0^1\,\hw\left(\dot\chi\,,\,\pi_1 \bfn \boldsymbol{f}_{a}\boldsymbol{f}^{a}\,{\cal G}(t)\right)\0\\
 =&\int_0^1\,\hw\left(\dot\chi\,,\,\pi_1 \bfn\, \bfU\,{\cal G}(t)\right),\0
 \end{align}
 where we have defined the higher order {\color{Black} odd} co-derivation associated to the open-closed Poisson bivector 
 \begin{align}
\bfU\coloneqq \frac{1}{2}(-)^{f^{a}}\boldsymbol{f}_{a}\boldsymbol{f}^{a},
 \end{align}
  which trivially obeys
  \begin{align}
  \bfU^2&=0,\\
  \pi_1\bfU&=0.
  \end{align}
  In total we have thus shown
  \begin{align}
  \frac12(S_{\rm oc},S_{\rm oc})+\Delta S_{\rm oc}=&\int_0^1\,\hw\left(\dot\chi\,,\,\pi_1 \bfn\,(\bfn+ \bfU)\,{\cal G}(t)\right)\0\\
  =&\int_0^1\,\hw\left(\dot\chi\,,\,\pi_1 (\bfn+ \bfU)^2\,{\cal G}(t)\right).
  \end{align}
  Therefore the BV quantum master equation is satisfied if the open-closed products assembled in the co-derivation $\bfn$ obey
  \begin{align}
  (\bfn+\bfU)^2=0.\label{nilpotent}
  \end{align}
  \subsection{Topological expansion}
  It is instructive to partially un-package this construction by making explicit the open and closed sectors. In this case the quantum BV master equation takes the form 
  \begin{equation}
\label{eqn: openclosedquantummaster}
    \begin{split}
       &\dfrac{1}{2}(S_{oc},S_{oc})+\Delta S_{oc} =\\
       &=\dfrac{1}{\kappa^{2}}\int_{0}^{1}dt\,\omega_{\rm c}\left(\dot\Phi,\pi_{10}\left\{\dfrac{1}{2}[\boldsymbol{l},\boldsymbol{l}]+\boldsymbol{l}\boldsymbol{m}+\left[\boldsymbol{l},\kappa\boldsymbol{U}_{o}+\kappa^{2}\boldsymbol{U}_{c}\right]\right\}{\cal G}(t)\right)+\\
       &+\dfrac{1}{\kappa^2}\int_{0}^{1}dt\,\omega_{\rm o}\left(\dot\Psi,\kappa\pi_{01}\left\{\dfrac{1}{2}[\boldsymbol{m},\boldsymbol{m}]+\boldsymbol{m}\boldsymbol{l}+\left[\boldsymbol{m},\kappa\boldsymbol{U}_{o}+\kappa^{2}\boldsymbol{U}_{c}\right]\right\}{\cal G}(t)\right),
    \end{split}
\end{equation}
where we have defined the higher order co-derivations associated to the closed and open Poisson bivectors
\begin{align}
\bfU&=\kappa^2\,  \boldsymbol{U}_{\rm c}+\kappa\, \boldsymbol{U}_{\rm o}\\
  \boldsymbol{U}_{\rm c}&=\dfrac{(-)^{c^{a}}}{2}\boldsymbol{c}^{a}\boldsymbol{c}_{a}\\
  \boldsymbol{U}_{\rm o}&=\dfrac{(-)^{o^{a}}}{2}\boldsymbol{o}^{a}\boldsymbol{o}_{a}.
\end{align}
We already know that the above expression is equal to zero if  the products satisfy the nilpotency condition \eqref{nilpotent} which now reads 
\begin{equation}
\label{eqn: quantumhomotopyalgebra}
    \left[ \boldsymbol{l}+\boldsymbol{m}+\kappa\boldsymbol{U}_{\rm o}+\kappa^{2}\boldsymbol{U}_{\rm c},\boldsymbol{l}+\boldsymbol{m}+\kappa\boldsymbol{U}_{\rm o}+\kappa^{2}\boldsymbol{U}_{\rm c}\right]=0.
\end{equation}
Now we would like to split  \eqref{eqn: quantumhomotopyalgebra} into the open and closed sectors by applying $\pi_{10}$ and $\pi_{01}$  and then expand in powers of $\kappa$. 
First of all we immediately notice that 
\begin{align}
 \pi_{10} \left[ \boldsymbol{l}+\boldsymbol{m}+\kappa\boldsymbol{U}_{\rm o}+\kappa^{2}\boldsymbol{U}_{\rm c},\boldsymbol{l}+\boldsymbol{m}+\kappa\boldsymbol{U}_{\rm o}+\kappa^{2}\boldsymbol{U}_{\rm c}\right]&=\pi_{10}\left\{\dfrac{1}{2}[\boldsymbol{l},\boldsymbol{l}]+\boldsymbol{l}\boldsymbol{m}+\left[\boldsymbol{l},\kappa\boldsymbol{U}_{o}+\kappa^{2}\boldsymbol{U}_{c}\right]\right\}\\
  \kappa\pi_{01} \left[ \boldsymbol{l}+\boldsymbol{m}+\kappa\boldsymbol{U}_{\rm o}+\kappa^{2}\boldsymbol{U}_{\rm c},\boldsymbol{l}+\boldsymbol{m}+\kappa\boldsymbol{U}_{\rm o}+\kappa^{2}\boldsymbol{U}_{\rm c}\right]&=\kappa\pi_{01}\left\{\dfrac{1}{2}[\boldsymbol{m},\boldsymbol{m}]+\boldsymbol{m}\boldsymbol{l}+\left[\boldsymbol{m},\kappa\boldsymbol{U}_{o}+\kappa^{2}\boldsymbol{U}_{c}\right]\right\}.
\end{align}
Notice that the factor of $\kappa$ in front of $\pi_{01}$ is necessary to work homogeneously in $\kappa$ and thus to get an homogeneous dependence on the topology of the involved Riemann surfaces when we switch from the open to closed symplectic forms and viceversa. If we now expand in $\kappa$ using (\ref{l-expand}, \ref{m-expand}) we get
\begin{itemize}
    \item $O(\kappa^{0})$: the $L_{\infty}$ relations controlling the consistency of closed string amplitudes on the sphere
    \begin{itemize} 
    \item $\kappa\pi_{01}$: \begin{equation}  \textrm{Nothing}.
        \end{equation}

      \item $\pi_{10}$:
        \begin{equation}  \dfrac{1}{2}[\boldsymbol{l}^{(0,0)},\boldsymbol{l}^{(0,0)}]=0.
        \end{equation}
        
    \end{itemize}
    \item $O(\kappa)$: These are  relations controlling the consistency of open-closed amplitudes on the disk, resulting in the SDHA of \cite{cosmo}
    \begin{itemize} 
     \item $\kappa\pi_{01}$: this sector deals with disk amplitudes with at least one open string and arbitrary number of closed strings. The reader can recognize Kajiura-Stasheff OCHA\footnote{\ch{Notice that here we had to compensate the removal of the projector $\pi_{01}$ from in front of the homotopy relation by formally turning the $\boldsymbol{m}^{(0,1)}\boldsymbol{l}^{(0,0)}$ term into a full graded commutator in order to remove the contributions where the multi-string products are not nested into each other. In this particular case this is an allowed operation because the $\boldsymbol{l}^{(0,0)}\boldsymbol{m}^{(0,1)}$ term would never contribute with nested products.}}  \cite{Kajiura:2004xu, Kajiura:2005sn}
        \begin{equation}
            \dfrac{1}{2}[\boldsymbol{m}^{(0,1)},\boldsymbol{m}^{(0,1)}]+\ch{[}\boldsymbol{m}^{(0,1)}\ch{,}\boldsymbol{l}^{(0,0)}\ch{]}=0.
        \end{equation}

        \item $\pi_{10}$: this sector is a dual version of the OCHA  (meaning no new constraints) when there is at least one open string input.  However it also contains the case with only closed strings on the disk which are not included in the OCHA
                \begin{equation}
[\boldsymbol{l}^{(0,0)},\boldsymbol{l}^{(0,1)}]+ \ch{\pi_{1,0}\,}\boldsymbol{l}^{(0,1)}\boldsymbol{m}^{(0,1)}=0,
        \end{equation}
          \end{itemize}
    \item $O(\kappa^{2})$: These are relations dealing with surfaces of vanishing Euler number: the torus (purely closed string amplitudes) and annulus (open-closed amplitudes)
   \begin{itemize} 
   \item $\kappa\pi_{01}$:
     \begin{equation}
         [\boldsymbol{m}^{(0,2)},\boldsymbol{m}^{(0,1)}]+ \ch{\pi_{0,1}\,}\boldsymbol{m}^{(0,1)}\boldsymbol{l}^{(0,1)}+\ch{[}\boldsymbol{m}^{(0,2)}\ch{,}\boldsymbol{l}^{(0,0)}\ch{]}+\ch{[}\boldsymbol{m}^{(0,1)}\ch{,}\boldsymbol{U}_{\rm o}\ch{]}=0.
     \end{equation}
     \item $\pi_{10}$:
     \begin{equation}
         \dfrac{1}{2}[\boldsymbol{l}^{(0,1)},\boldsymbol{l}^{(0,1)}]+[\boldsymbol{l}^{(0,0)},\boldsymbol{l}^{(0,2)}+\boldsymbol{l}^{(1,0)}]+\ch{[}\boldsymbol{l}^{(0,1)}\ch{,}\boldsymbol{U}_{\rm o}\ch{]}+\ch{[}\boldsymbol{l}^{(0,0)}\ch{,}\boldsymbol{U}_{\rm c}\ch{]}=0,
     \end{equation}
    
  \end{itemize}
  \item $O(\kappa^3)$: Surfaces of Euler number $\chi=-1$, and so on...
\end{itemize}
Thus in total, paying attention that the open and closed symplectic forms are normalized with different powers of $\kappa$, we see that the nilpotency relation \eqref{nilpotent} has a precise perturbative expansion in $\kappa$ which correctly follows the expansion in the Euler number. Notice however that this expansion is not a real topological expansion because different topologies can have the same Euler number (for example the torus and the annulus). In the companion paper \cite{large-N} we will fully resolve the topology  by formulating OC-SFT on a set of $N$ identical D-branes.

    \section{Conclusions}\label{sec:5}
    
     The main point of this paper is  that the final algebraic structure of open-closed string field theory is that of a loop homotopy algebra \cite{Markl:1997bj}, the same algebraic structure governing quantum closed string field theory, but defined on a more complicated space where both closed strings and colour ordered open strings can be fit. We expect this formulation to have many advantages.

  Starting from our microscopic BV description it would be  interesting to algebraically construct the 1PI effective action \cite{Moosavian:2019ydz, Sen:2014dqa} and verify that it will be described by an open-closed co-derivation  $\bfn^{(1PI)}$ satisfying $\left(\bfn^{(1PI)}\right)^2=0$, in order to solve the classical BV master equation.  This will provide an example of a ``classical'' open closed nilpotent structure and it would be interesting to use this 1PI framework to address quantum background independence in open-closed SFT, extending the analysis of \cite{cosmo} to quantum corrections.

  Another interesting application of this new formulation is the possibility of directly using the  homotopy transfer to perform the open-closed target space path integral and integrate out different sectors of the theory, thus obtaining different effective descriptions of the same physics, depending on which degrees of freedom have been integrated out.  By choosing a projector $P$ acting on ${\cal H}_{\rm o}\oplus{\cal H}_{\rm c}$ projecting on the states we want to retain, we could use the (quantum) homotopy transfer \cite{Sen:2016qap, Erbin:2020eyc, Koyama:2020qfb, Okawa:2022sjf} by constructing an appropriate propagator $h$ such that
    \begin{align}
 [Q_{\rm open}+   Q_{\rm closed}, \,h]=\boldsymbol{1}_{{\cal H}_{\rm o}\oplus{\cal H}_{\rm c}}-P,
    \end{align}
    so that the effective action for the fields in the image of $P$ will be governed by a theory which will have the same structure as the initial OC SFT with a new nilpotent structure which will be given by
    \begin{align}
\tilde  \bfn+\tilde \bfU=&P\,(\bfn+\bfU)\frac1{1+\boldsymbol{h}(\bfn-\boldsymbol{Q}+\bfU)}P\\
(\tilde  \bfn+\tilde \bfU)^2=&0.
    \end{align}
    It is natural to wonder about the structure of the effective action which can be obtained integrating out completely open or closed strings by choosing the corresponding projectors that keep either ${\cal H}_{\rm o}$ or ${\cal H}_{\rm c}$ out of the full ${\cal H}_{\rm o}\oplus{\cal H}_{\rm c}$.    This seems a promising QFT framework to understand microscopically  the gauge/gravity duality and geometric transitions in general.  A preliminary analysis in this direction is presented in \cite{large-N}.

Another interesting direction in which the present formulation is expected to be useful is the discussion of the ghost-dilaton theorem  \cite{Bergman:1994qq, Belopolsky:1995vi, Erler:2022agw} in open-closed SFT \cite{progress1}.

Of course one would ultimately want to \ch{generalize this discussion to} the context of the superstring. It should be possible  to extend our construction to Type II superstrings, starting from the initial analysis of the superstring SDHA \cite{super-SDHA}, where the construction of the open-closed vertices on the disk has been discussed.

    \section*{Acknowledgments}
     JV thanks INFN Turin for their hospitality during the initial stages of this work.     The work of CM  and AR  is partially supported by the MUR PRIN contract 2020KR4KN2 “String Theory as a bridge between Gauge Theories and Quantum Gravity” and by the INFN project ST$\&$FI “String Theory and Fundamental Interactions”.    The work of JV was supported by the NCCR SwissMAP that
    is funded by the Swiss National Science Foundation.

    \appendix
    
\section{Cyclicity and 0-product co-derivations}\label{app1}

In this appendix we  show that 
\begin{equation}
\label{eqn: cyclicitystatement}
\hat{\omega}\left(f^{a},\pi_{1}\boldsymbol{n} \boldsymbol{f}_{a}\cal{G}\right)=(-)^{f^{a}}\int_{0}^{1}dt\,\hat{\omega}\left(\dot{\chi}(t), \pi_{1}\boldsymbol{n}\boldsymbol{f_{a}}\boldsymbol{f^{a}}\mathcal{G}(t)\right)=2\int_{0}^{1}dt\,\hat{\omega}\left(\dot{\chi}(t), \pi_{1}\boldsymbol{n}\boldsymbol{U}\mathcal{G}(t)\right).
\end{equation}
First of all, by unpackaging the l.h.s. of the above equation we get
\begin{equation}
\label{eqn: split}
\hat{\omega}\left(f^{a},\pi_{1}\boldsymbol{n} \boldsymbol{f}_{a}\cal{G}\right)=\wc\left(c^{a},\pi_{10}\boldsymbol{l}\boldsymbol{c}_{a}\mathcal{G}\right)+\wo\left(o^{a},\pi_{01}\boldsymbol{m}\boldsymbol{o}_{a}\mathcal{G}\right).
\end{equation}
Now we proceed by manipulating the two terms of ~\eqref{eqn: split} separately 
\begin{equation}
\label{eqn: cc}
    \begin{split}
&\wc\left(c^{a},\pi_{10}\boldsymbol{l}\boldsymbol{c}_{a}\mathcal{G}\right)\ch{=} \\
&\ch{=}
\sum_{g,b,k,\{p_1,...,p_b\}\ch{=0}}^{\infty}\dfrac{\kappa^{2g+b}}{b!k!(p_1)\cdot \cdot\cdot (p_b)}\wc\left(c^{a},l_{k+1[p_1,...,p_b]}^{(g,b)}\left(c_{a}\wedge \Phi^{\wedge k}\otimes' \Psi^{\odot p_1} \cdots  \Psi^{\odot p_b}\right)\right)\\
&=\
\sum_{g,b,\{p_1,...,p_b\}\ch{=0},k\ch{=1}}^{\infty}\dfrac{(-)^{c^{a}}\kappa^{2g+b}}{b!k!(p_1)\cdot \cdot\cdot (p_b)}\wc\left(\Phi,l_{k+1[p_1,...,p_b]}^{(g,b)}\left(c^{a}\wedge c_{a}\wedge \Phi^{\wedge k-1}\otimes' \Psi^{\odot p_1} \cdots \Psi^{\odot p_b}\right)\right)\nonumber\\
&\ch{+\sum_{g,b,\{p_1,...,p_b\}\ch{=0}}^{\infty}\dfrac{\kappa^{2g+b}}{b!(p_1)\cdot \cdot\cdot (p_b)}\wc\left(c^{a},l_{1[p_1,...,p_b]}^{(g,b)}\left(c_{a}\otimes' \Psi^{\odot p_1} \cdots  \Psi^{\odot p_b}\right)\right)}\\
&\ch{=\int_{0}^{1}dt\sum_{g,b,\{p_1,...,p_b\}\ch{=0},k=1}^{\infty}\dfrac{(-)^{c^{a}}\kappa^{2g+b}}{b!(k-1)!(p_1)\cdots  (p_b)}\wc\left(\dot{\Phi},l_{k+1[p_1,...,p_b]}^{(g,b)}\left(c^{a}\wedge c_{a}\wedge \Phi^{\wedge k-1}\otimes' \Psi^{\odot p_1} \cdots \Psi^{\odot p_b}\right)\right)+}\\
&\ch{+\int_{0}^{1}dt\sum_{g,b,\{p_1,...,p_b\}\ch{=0},k=1}^{\infty}\dfrac{(-)^{c^{a}}\kappa^{2g+b}}{b!k!(p_1)\cdot \cdot\cdot (p_b)}\wc\left(\Phi,l_{k+1[p_1,...,p_b]}^{(g,b)}\left(c^{a}\wedge c_{a}\wedge \Phi^{\wedge k-1}\otimes' \p_t(\Psi^{\odot p_1} \cdots \Psi^{\odot p_b})\right)\right)}\nonumber\\
&\ch{+\int_{0}^{1}dt\sum_{g,b,\{p_1,...,p_b\}\ch{=0}}^{\infty}\dfrac{\kappa^{2g+b}}{b!(p_1)\cdot \cdot\cdot (p_b)}\wc\left(c^{a},l_{1[p_1,...,p_b]}^{(g,b)}\left(c_{a}\otimes' \p_t(\Psi^{\odot p_1} \cdots  \Psi^{\odot p_b})\right)\right)}
    \end{split}
\end{equation}
\ch{where we have noticed, that the contribution containing zero closed strings needs to be treated separately. Here we note that we can first rewrite the two terms where the derivative acts on the open string fields as}
\begin{align}
&\ch{\int_{0}^{1}dt\sum_{\substack{g,b,\{p_2,...,p_b\}\ch{=0},k=1\\ p_1=1}}^{\infty}\dfrac{(-)^{c^{a}}\kappa^{2g+b}}{b!k!(p_1)\cdot \cdot\cdot (p_b)}\wc\left(\Phi,l_{k+1[p_1,...,p_b]}^{(g,b)}\left(c^{a}\wedge c_{a}\wedge \Phi^{\wedge k-1}\otimes' p_1\dot{\Psi}\odot\Psi^{\odot p_1-1} \cdots \Psi^{\odot p_b}\right)\right)+\ldots}\nonumber\\
&\ch{\ldots+\int_{0}^{1}dt\sum_{\substack{g,b,\{p_1,...,p_{b-1}\}\ch{=0},k=1\\ p_b=1}}^{\infty}\dfrac{(-)^{c^{a}}\kappa^{2g+b}}{b!k!(p_1)\cdot \cdot\cdot (p_b)}\wc\left(\Phi,l_{k+1[p_1,...,p_b]}^{(g,b)}\left(c^{a}\wedge c_{a}\wedge \Phi^{\wedge k-1}\otimes' \Psi^{\odot p_1} \cdots p_b\dot{\Psi}\odot\Psi^{\odot p_b-1}\right)\right)}\,,
\end{align}
and 
\begin{align}
	&\ch{\int_{0}^{1}dt\sum_{\substack{g,b,\{p_2,...,p_b\}\ch{=0}\\
		p_1=1
	}}^{\infty}\dfrac{\kappa^{2g+b}}{b!(p_1)\cdot \cdot\cdot (p_b)}\wc\left(c^{a},l_{1[p_1,...,p_b]}^{(g,b)}\left(c_{a}\otimes' p_1\dot{\Psi}\odot\Psi^{\odot p_1-1} \cdots  \Psi^{\odot p_b}\right)\right)+\ldots}\nonumber\\
	&\ch{\ldots+\int_{0}^{1}dt\sum_{\substack{g,b,\{p_1,...,p_{b-1}\}\ch{=0}\\
		p_b=1		
		}}^{\infty}\dfrac{\kappa^{2g+b}}{b!(p_1)\cdot \cdot\cdot (p_b)}\wc\left(c^{a},l_{1[p_1,...,p_b]}^{(g,b)}\left(c_{a}\otimes' \Psi^{\odot p_1} \cdots  p_b\dot{\Psi}\odot\Psi^{\odot p_b-1}\right)\right)}\,.
	\end{align}
\ch{Using the open-closed duality and relabelling the summation indices, the two terms combine into}
\begin{align}
	&\ch{\int_{0}^{1}dt\sum_{\substack{g,b,\{p_1,...,p_{b-1}\}\ch{=0},k=0\\ p_b=1}}^{\infty}\dfrac{(-)^{c^{a}}\kappa^{2g+b-1}bp_b}{b!k!(p_1)\cdot \cdot\cdot (p_b)}\kappa\wo\left(\dot{\Psi},m_{k+2[p_1,...,p_{b-1}]p_b-1}^{(g,b)}\left(c^{a}\wedge c_{a}\wedge \Phi^{\wedge k}\otimes' \Psi^{\odot p_1} \cdots \Psi^{\odot p_{b-1}} \otimes'' \Psi^{\otimes p_b-1}\right)\right)\,.}
\end{align}
We can then continue to manipulate
\begin{equation}
	\begin{split}
		&\wc\left(c^{a},\pi_{10}\boldsymbol{l}\boldsymbol{c}_{a}\mathcal{G}\right)\ch{=} \\
&=\int_{0}^{1}dt\sum_{g,b,\{p_1,...,p_b\}\ch{=0},\ch{k=1}}^{\infty}\dfrac{(-)^{c^{a}}\kappa^{2g+b}}{b!(k-1)!(p_1)\cdots  (p_b)}\wc\left(\dot{\Phi},l_{k+1[p_1,...,p_b]}^{(g,b)}\left(c^{a}\wedge c_{a}\wedge \Phi^{\wedge k-1}\otimes' \Psi^{\odot p_1} \cdots \Psi^{\odot p_b}\right)\right)+\\
&+\int_{0}^{1}dt\ch{\sum_{\substack{g,b,k,\{p_1,...,p_{b-1}\}=0\\ p_b=1}}^{\infty}}\dfrac{(-)^{c^{a}}\kappa^{2g+b}bp_b}{b!k!(p_1)\cdots  (p_b)}\,\kappa\wo\left(\dot{\Psi},m_{k+2[p_1,...,p_{b-1}]p_b-1}^{(g,b)}\left(c^{a}\wedge c_{a}\wedge \Phi^{\wedge k}\otimes'\Psi^{\odot p_1} \cdots  \Psi^{\odot p_{b-1}}\otimes'' \Psi^{\otimes p_b-1}\right)\right)\\
&\ch{=\int_{0}^{1}dt\sum_{g,b,\{p_1,...,p_b\}\ch{=0},k=1}^{\infty}\dfrac{(-)^{c^{a}}\kappa^{2g+b}}{b!(k-1)!(p_1)\cdots  (p_b)}\wc\left(\dot{\Phi},l_{k+1[p_1,...,p_b]}^{(g,b)}\left(c^{a}\wedge c_{a}\wedge \Phi^{\wedge k-1}\otimes' \Psi^{\odot p_1} \cdots \Psi^{\odot p_b}\right)\right)+}\\
&\ch{+\int_{0}^{1}dt\ch{\sum_{{g,b,k,\{p_1,...,p_b\}=0 }}^{\infty}}\dfrac{(-)^{c^{a}}\kappa^{2g+b}\ch{b(p_b)}}{b!k!(p_1)\cdots  (p_b)}\,\kappa\wo\left(\dot{\Psi},m_{k+2[p_1,...,p_{b-1}]p_b}^{(g,b)}\left(c^{a}\wedge c_{a}\wedge \Phi^{\wedge k}\otimes'\Psi^{\odot p_1} \cdots  \Psi^{\odot p_{b-1}}\otimes'' \Psi^{\otimes p_b}\right)\right)}\\
&=(-)^{c^{a}}\int_{0}^{1}dt\left\{\wc\left(\dot{\Phi}(t),\pi_{10}\boldsymbol{l}\boldsymbol{c}_{a}\boldsymbol{c}^{a}\mathcal{G}(t)\right)+\kappa\wo\left(\dot{\Psi}(t),\pi_{01}\boldsymbol{m}\boldsymbol{c}_{a}\boldsymbol{c}^{a}\mathcal{G}(t)\right)\right\}\\
&=(-)^{c^{a}}\int_{0}^{1}dt\,\hat{\omega}\left(\dot{\chi}(t),\pi_{1}\boldsymbol{n}\left(\kappa^{2}\boldsymbol{c}_{a}\boldsymbol{c}^{a}\right)\mathcal{G}(t)\right).
    \end{split}
\end{equation}
As for the open part we have that 
\begin{equation}
    \begin{split}
&\wo\left(o^{a},\pi_{01}\boldsymbol{m}\boldsymbol{o}_{a}\mathcal{G}\right)=\\    
&=\sum_{g,b,k,\{p_1,...,p_b\}}^{\infty}\dfrac{(\kappa^{2g+b-1})\ch{(p_1)}b(b-1)\ch{(p_b)}}{b!k!(p_1)\cdot \cdot\cdot (p_b)}\wo\left(o^{a},m^{(g,b)}_{k[p_1+1,..., p_{b-1}]p_b}\left(
        \Phi^{\wedge k}\otimes' o_{a}\odot \Psi^{\odot p_1} \cdots  \Psi^{\odot p_{b-1}}\otimes''\Psi^{\otimes p_b}\right)\right)+\\
        &+\sum_{g,b,k,\{p_1,...,p_b\}}^{\infty}\dfrac{(\kappa^{2g+b-1})\ch{(p_1)}b}{b!k!(p_1)\cdots (p_b)}\wo\left(o^{a},m^{(g,b)}_{k[p_b,...,p_{b-1}]p_1+1}\left(
        \Phi^{\wedge k}\otimes' \Psi^{\odot p_b} \cdots  \Psi^{\odot p_{b-1}} \otimes'' o_{a}\odot \Psi^{\odot p_1}\right)\right)\,.
    \end{split}
\end{equation}
\ch{By} using cyclicity, we can bring the basis vectors within the $\boldsymbol{m}$ product and relabeling the summation indices we get \ch{(also isolating the terms where this is not possible, because there are no open-string fields to be brought on the bra of the symplectic form -- those will eventually end up contributing into the $\bs{l}$-part of the final result \eqref{eqn: oo})}
\begin{equation}
    \begin{split}
 &\wo\left(o^{a},\pi_{01}\boldsymbol{m}\boldsymbol{o}_{a}\mathcal{G}\right)=\\   
        &= \sum_{\ch{g},b,k,\{p_3,...,p_b\}}^{\infty}\dfrac{(\kappa^{2g+b-1})(-)^{o^{a}}b(b-1)(p_1-1)(p_2-1)}{b!k!(p_1-1)(p_2-1)p_3\cdots  (p_b+1)}\ch{\times}\\&\qquad\qquad\qquad\qquad\ch{\times}\wo\left(\Psi,m^{(g,b)}_{k[p_1,...,p_{b-1}]p_b}\left(
        \Phi^{\wedge k}\otimes' o_{a}\odot \Psi^{\odot p_1-1}\wedge'o^{a}\odot \Psi^{\odot p_2-1} \cdots  \Psi^{\odot p_{b-1}}\otimes''\Psi^{\otimes p_b}\right)\right)+\\
        &+\sum_{\ch{g},b,k,\{p_1,...,p_b\}}^{\infty}\dfrac{(\kappa^{2g+b-1})(-)^{o^{a}}b(p_1-2)}{b!k!(p_1-2)p_2\cdots (p_b+1)}\ch{\times}\\&\qquad\qquad\ch{\times}\wo\left(\Psi,m^{(g,b)}_{k[p_1,...,p_{b-1}]p_b}\left(
        \Phi^{\wedge k}\otimes'\left[ \sum_{n=0}^{p_1-2} o^{a}\odot\Psi^{\odot n} \odot o_{a}\odot \Psi^{\odot p_1-n-2}   \right]  \cdots \Psi^{\odot p_{b-1}} \otimes''\Psi^{\otimes p_b}\right)\right)+\nonumber\\
        &\ch{+\sum_{g,b,k}^{\infty}\dfrac{(\kappa^{2g+b-1})b(b-1)}{b!k!}\wo\left(o^{a},m^{(g,b)}_{k[1,..., 0]0}\left(
        \Phi^{\wedge k}\otimes' o_{a} \otimes''\right)\right)+}\\
        &\ch{+\sum_{g,b,k}^{\infty}\dfrac{(\kappa^{2g+b-1})b}{b!k!}\wo\left(o^{a},m^{(g,b)}_{k[0,...,0]1}\left(
        \Phi^{\wedge k} \otimes'' o_{a}\right)\right)\,.}
    \end{split}
\end{equation}
At this point the meaning of the relationship written above is clear, i.e., the open BV Laplacian  glues a Riemann surface in two ways either by joining two punctures from the same boundary or from different boundaries, which correspond respectively to $(\cdots o^{a}\odot o_{a}\odot\Psi^{\odot p_1}\cdots)$ and $(\cdots o^{a}\odot \Psi^{\odot p_1}\wedge'  o_{a}\odot \Psi^{\odot p_2}\cdots)$. Now as discussed for the closed part we want to return to the co-algebra formalism by reconstructing the group like elements. In order to do so we apply $\int_{0}^{1} dt\,\partial_{t}$ isolating in the first slot of the symplectic form $\dot{\Psi}$ and $\dot{\Phi}$. After some algebraic steps the previous expression becomes
\begin{equation}
\label{eqn: oo}
    \begin{split}
&\wo\left(o^{a},\pi_{01}\boldsymbol{m}\boldsymbol{o}_{a}\mathcal{G}\right)=\\   
&= (-)^{o^{a}}\int_{0}^{1}dt\left\{\wo\left(\dot{\Psi}(t),\pi_{01}\boldsymbol{m}\boldsymbol{o}_{a}\boldsymbol{o}^{a}
\mathcal{G}(t)\right)+\kappa^{-1}\wc\left(\dot{\Phi}(t),\pi_{10}\boldsymbol{l}\boldsymbol{o}_{a}\boldsymbol{o}^{a}
\mathcal{G}(t)\right)\right\}=\\
&=(-)^{o^{a}}\int_{0}^{1}dt\,\hat{\omega}\left(\dot{\chi}(t),\pi_{1}\boldsymbol{n}\left(\kappa\boldsymbol{o}_{a}\boldsymbol{o}^{a}\right)\mathcal{G}(t)\right).
    \end{split}
\end{equation}
Furthermore, to obtain the previous result, it is essential to observe that two open 0-coderivations on the group like element act as follows 
\begin{equation}
\begin{split}
&\boldsymbol{o}_{a}\boldsymbol{o}^{a}\mathcal{G} =\\
     &= \sum_{b,k,\{p_1,...,p_b\}}^{\infty}\dfrac{b(b-1)\ch{(p_1)(p_2)}}{b!k!(p_1)\cdot \cdot\cdot (p_b)}\left(\Phi^{\wedge k} \otimes' o_{a}\odot\Psi^{\odot p_1}\wedge' o^{a}\odot \Psi^{p_2} \cdots\Psi^{\odot p_b}\right)+\\
     &+ \sum_{b,k,\{p_1,...,p_b\}}^{\infty}\dfrac{b\ch{(p_1)}}{b!k!(p_1)\cdot \cdot\cdot (p_b)}\left(
        \Phi^{\wedge k}\otimes'\left[ \sum_{n=0}^{p_1} o^{a}\odot\Psi^{\odot n} \odot o_{a}\odot \Psi^{\odot p_1-n}   \right] \cdots  \Psi^{\odot p_b}\right).
\end{split}
\end{equation}
Finally by adding ~\eqref{eqn: cc} with ~\eqref{eqn: oo} we get the desired result
\begin{equation}
    \begin{split}
\hat{\omega}\left(f^{a},\pi_{1}\boldsymbol{n} \boldsymbol{f}_{a}\cal{G}\right)&=\int_{0}^{1}dt\,\hat{\omega}\left(\dot{\chi}(t),\pi_{1}\boldsymbol{n}\left((-)^{o^{a}}\kappa\boldsymbol{o}_{a}\boldsymbol{o}^{a}+(-)^{c^{a}}\kappa^{2}\boldsymbol{c}_{a}\boldsymbol{c}^{a}\right)\mathcal{G}(t)\right)=\\
&=(-)^{f^{a}}\int_{0}^{1}dt\,\hat{\omega}\left(\dot{\chi}(t), \pi_{1}\boldsymbol{n}\boldsymbol{f_{a}}\boldsymbol{f^{a}}\mathcal{G}(t)\right).
    \end{split}
\end{equation}

    \endgroup
    \end{document}